\documentclass[pra,aps,showpacs,groupedaddress,superscriptaddress,twocolumn]{revtex4-1}
\usepackage{amsmath,amssymb}
\usepackage[utf8x]{inputenc}
\usepackage[english]{babel}
\usepackage{graphicx,epsfig}
\usepackage{todonotes}
\usepackage{hyperref}

\renewcommand{\vec}{\mathbf}

 \global\long\def\abs#1{\left|#1\right|}

  \global\long\def\v#1{\vec{#1}}

\usepackage{braket}
\usepackage{soul}
\usepackage{color}
\makeatother
\usepackage{placeins}

\begin{document}

\title{Quantum Cherenkov transition of finite momentum Bose polarons}

\author{Kushal Seetharam}
\affiliation{Department of Electrical Engineering, Massachusetts Institute of Technologies, Cambridge, Massachusetts 02139, USA}
\affiliation{Department of Physics, Harvard University, Cambridge, Massachusetts 02138, USA}

\author{Yulia Shchadilova}
\affiliation{Department of Physics, Harvard University, Cambridge, Massachusetts 02138, USA}

\author{Fabian Grusdt}
\affiliation{Department of Physics and Arnold Sommerfeld Center for Theoretical Physics (ASC), Ludwig-Maximilians-Universit\"at M\"unchen, Theresienstr. 37, M\"unchen D-80333, Germany}
\affiliation{Munich Center for Quantum Science and Technology (MCQST), Schellingstr. 4, D-80799 M\"unchen, Germany}

\author{Mikhail Zvonarev}
\affiliation{Universit\'e Paris-Saclay, CNRS, LPTMS, 91405, Orsay, France}
\affiliation{Russian Quantum Center, Skolkovo, Moscow 143025, Russia}
\affiliation{St. Petersburg Department of V.A. Steklov Mathematical Institute of Russian Academy of Sciences, Fontanka 27, St. Petersburg, 191023, Russia}

\author{Eugene Demler}
\affiliation{Department of Physics, Harvard University, Cambridge, Massachusetts 02138, USA}
\affiliation{Institute for Theoretical Physics, ETH Z{\"u}rich, 8093 Z{\"u}rich, Switzerland}

\date{\today}

\begin{abstract}
We investigate the behavior of a finite-momentum impurity immersed in a weakly interacting three-dimensional Bose-Einstein condensate (BEC) of ultra-cold atoms, giving a detailed account of the dynamical quantum Cherenkov transition discussed in Ref.~\cite{Seetharam_CherenkovPRL2021}. Using a time-dependent variational approach, we identify a transition in the far-from-equilibrium dynamics of the system after the attractive short-range impurity-boson interaction is quenched on. The transition occurs as the impurity's velocity crosses an interaction-dependent critical value, and manifests in the long-time behavior of the Loschmidt echo and average impurity velocity. This behavior is also reflected in the finite momentum ground state of the system, where the group velocity of the interaction-dressed impurity loses it's dependence on the total momentum of the system as the critical point is crossed. The transition we discuss should be experimentally observable via a variety of common protocols in ultracold atomic systems such as time-of-flight imaging, RF spectroscopy, Ramsey interferometry, and absorption imaging.

\end{abstract}


\newpage

\maketitle

\section{Introduction}
\label{sec:Intro}
The non-equilibrium dynamics of quantum many-body systems is amongst
the most actively researched areas of modern physics. One of the most
prototypical problems that underlies the physics behind several such
systems is that of the quantum impurity, where an impurity degree
of freedom interacts with a surrounding many-body system. The equilibrium state of this setting is often well-described by the notion of a polaron
quasiparticle, where the impurity is dressed by excitations of the surrounding bath which renormalize its mass, energy, and other properties~\cite{Landau1933,Pekar1946,Landau1948,Frohlich1954}. As polarons are an archetypal quasiparticle, investigating the connection between their ground state properties and their far-from-equilibrium dynamics may help elucidate the non-equilibrium behavior of quasiparticles which characterize other quantum many-body systems~\cite{Skou2021,Gribben2024}.

In modern times, ultracold atomic gases have provided an experimental
platform that allow a much more detailed probing of polaronic physics \cite{Mathey2004,Grusdt2024}.
Control of host atom species allows specification of fermionic or
bosonic statistics and effective spin degrees of freedom. Additionally,
tunability of the interaction strength between the impurity and host
atom through Feshbach resonances and powerful measurement techniques
including radio frequency (RF) spectroscopy, Ramsey interferometery, absorption imaging, and time-of-flight imaging
enable a detailed study of these systems.

Experiments studying impurities in ultracold gases have primarily
examined their RF spectra~\cite{Hu2016,Jorgensen2016,Yan2019}, with recent experiments starting to probe their non-equilibrium dynamics~\cite{Skou2021} and collective behavior~\cite{Yan2024}.
Theoretically, the equilibrium Bose polaron has been well-studied
and there have recently been several works on dynamics as well. Variational
methods have been used to predict RF spectra, average values of different
observables, spatial density profiles, and even systems with multiple
impurities \cite{Shchadilova2016,VanLoon2018,Drescher2019,Shchadilova2016a,Shashi2014,SantiagoGarcia2024}.
T-matrix approximations have also been used to study the RF spectra
while confirming the importance of beyond-Fr\"ohlich terms in the Hamiltonian
at strong interactions \cite{Rath2013}. Renormalization group methods
have been used to establish the stability of the ground state quasiparticle~\cite{Hinrichs2023} and examine trajectories and the effect of quantum fluctuations
on top of the mean-field solution \cite{Grusdt2015,Grusdt2017,Grusdt2017a,Grusdt2018}.
Markovian master equations have been used to study thermalization
dynamics and polaron formation \cite{Lausch2018,Nielsen2019}, while
finite temperature effects have also been studied with diagrammatic
techniques for strong coupling \cite{Guenther2018} and perturbation
theory for weak coupling \cite{Levinsen2017}. Non-perturbative methods have also been used to examine impurity dynamics in harmonically trapped 1D BECs  \cite{Mistakidis2019, Mistakidis2019a} and the subsonic-supersonic dynamics of weakly interacting impurities \cite{Boyanovsky2019}. Quantum Monte Carlo computations have also been used to study the energy and other properties of the Bose polaron \cite{Ardila2015,Ardila2019}. Efforts to study the strong coupling regime by moving beyond Bogoliubov theory have been made \cite{Drescher2020}, and the collective behavior of many impurities immersed in a BEC has been studied using a Boltzmann analysis~\cite{Dolgriev2024}.


Previous works using variational states primarily studied the system at zero momentum~\cite{Shchadilova2016, Drescher2019} or its dynamics at specific values of finite momentum~\cite{VanLoon2018}. Additionally, some theoretical
work has previously been done on fast impurities immersed in a BEC, including BECs flowing supersonically across an infinite-mass
defect  \cite{Carusotto2006}, supersonic impurities in a 1D quantum
liquid \cite{Mathy2012}, and a semiclassical treatment of weakly-interacting
supersonic impurities in a 3D BEC \cite{Dasenbrook2013}.


\begin{figure*}[t!]
	\centering
	\includegraphics[width=0.99\textwidth]{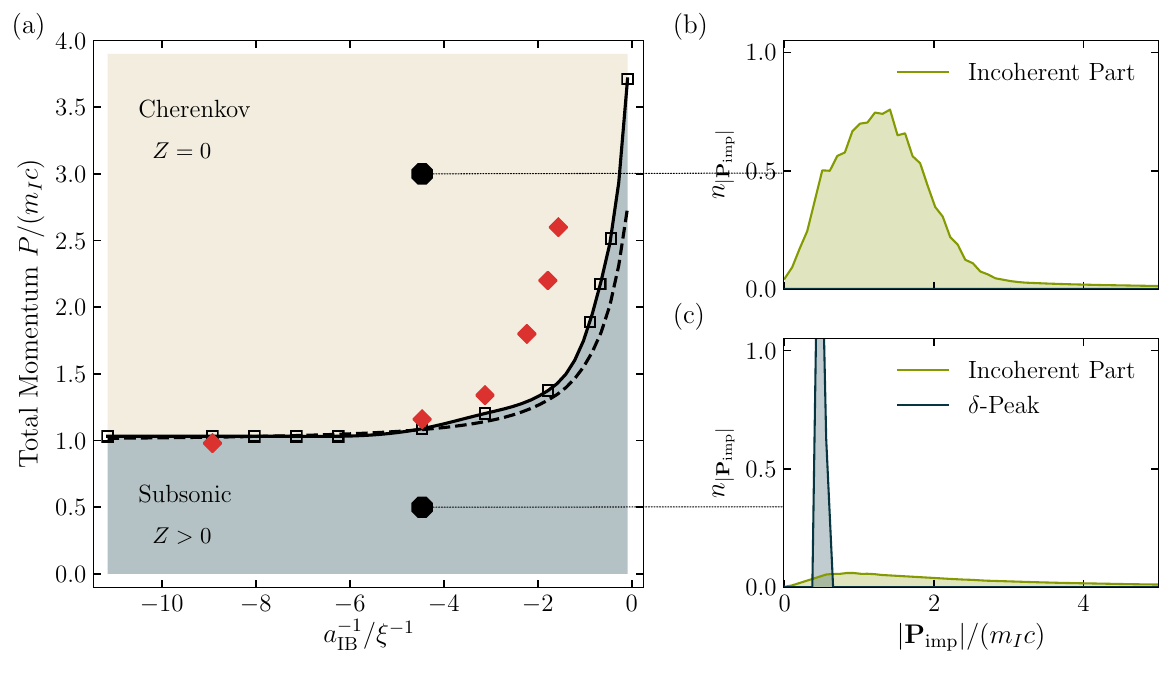}
	\caption{Quantum Cherenkov transition of a finite momentum impurity interacting with a weakly-interacting 3D BEC~\cite{Seetharam_CherenkovPRL2021}. (a) Phase diagram. The black solid line marks the numerically extracted critical momentum from the discontinuity in the second derivative of the FMGS energy. The black dashed line marks the predicted transition value $m^{*}c$. The critical momentum describing the transition between subsonic and Cherenkov regimes increases with interaction strength and is related to polaronic mass enhancement. The red diamonds represent the numerically extracted transition points in the long time limit of the quench dynamics. Panels (b) and (c) illustrate the impurity's momentum magnitude distribution in each regime. Note that we have introduced Gaussian broadening of the $\delta$-peak in the distributions by hand; in reality we get a $\delta$-peak at $|\mathbf{P}_{\rm imp}|=P$ with a weight $Z$. The mass ratio is $m_{I}/m_{B}=1$ and we use a momentum space UV cutoff $\Lambda=9.27[\xi^{-1}]$ where $\xi$ is the healing length of the BEC.}
	\label{fig:GS_PhaseDiagram}
\end{figure*}
 
In this paper, we study a finite momentum impurity immersed in
a 3D Bose-Einstein condensate (BEC) using variational states that go beyond the semiclassical treatment and allow study of a wide range of interaction strengths. Specifically, we focus on the case of an effective attractive interaction between the impurity and the BEC, thoroughly exploring different initial impurity momentum values and giving a detailed account of the quantum Cherenkov transition this system exhibits~\cite{Seetharam_CherenkovPRL2021}. First, we examine the finite momentum ground state (FMGS) of the system, which is the lowest energy state of the system at a finite value of the conserved total momentum. Second, we study dynamics when the impurity is quenched from a non-interacting
state to an interacting state. While the system's state after a quench is not guaranteed to approach the true FMGS, the long-time behavior of certain observables are still related to their FMGS values, thus connecting the physics of the two scenarios. 
In both the FMGS and quench dynamics, we find two qualitatively different regimes of the system's behavior as the total system momentum is increased. At low momenta, the system behaves in accordance with the typical picture of a polaron quasiparticle; the FMGS has a well-defined effective mass associated with a quadratic energy-momentum relation and both the FMGS and infinite time state during quench dynamics are connected to the free impurity. At high enough momenta, the impurity's average velocity equals the speed of sound of the low-lying excitations of the bosonic bath even after interactions between the impurity and BEC redistribute momenta into Bogoliubov excitations of the bath. In this `Cherenkov' regime, the interaction-dressed impurity still has a well-defined energy-momentum relation in the FMGS, but this relation is linear rather than quadratic and the polaron is no longer connected to the free impurity. 


The paper is organized as follows. We start by giving a summary of results in Sec. \ref{sec:SumRes}. Next, we introduce the system in Sec. \ref{sec:System}. We discuss the variational method used to derive equations of motion for a single impurity in a BEC. In Sec. \ref{sec:GroundState}, we study the Cherenkov transition in the FMGS using imaginary time dynamics of the equations of motion. In Sec. \ref{sec:ReDyn}, we examine manifestations of the transition in real-time dynamics. We close with a conclusion in Sec. \ref{sec:Conclusion}.

\section{Summary of Results}
\label{sec:SumRes}

We start with a closed system consisting of an impurity immersed in a BEC where the total momentum of the system is conserved. We first study the FMGS properties of the system and then examine quench dynamics where an initially free impurity is instantaneously made to interact with the atoms comprising the BEC around it. The effective contact potential describing this interaction can be either attractive or repulsive. In this work, we focus on the regime of effective attractive interactions between the impurity and the BEC. To theoretically treat the system, we first entangle the impurity and BEC degrees of freedom using a non-Gaussian transformation. Such a transformation makes it possible to capture non-trivial correlations between the impurity and the BEC using a simple coherent state variational ansatz for just the BEC degrees of freedom. The time-dependent variational principle can then be used to derive equations of motion for the system for arbitrary interaction strengths and mass ratios.

(I) \textit{FMGS Transition}. 
If we pick an interaction strength and calculate the FMGS energy as a function of total momentum $\v P$, we find that this relation is initially quadratic and then becomes linear at a critical momentum $\vec{P}_\textrm{crit}$. We posit that the energy-momentum relation of a finite momentum impurity immersed in a weakly interacting BEC has the form
\begin{equation}\label{eq:PolDispersion}
	E\left(\vec P\right) = \begin{cases}
	\frac{\vec{P}^{2}}{2m^{*}}, & \text{if $\abs{\v P}\leq\abs{\v P_{\rm crit}}$}\\
	E_c + c\abs{\vec{P}}, & \text{if $\abs{\v P}>\abs{\v P_{\rm crit}}$}
	\end{cases}
\end{equation}
where
\begin{align}\label{eq:Pcrit}
\abs{\v P_{\rm crit}}=m^{*} c.
\end{align}
Note that the dispersion is continuous at all momenta and $E_c=c\abs{\vec{P}_\textrm{crit}}/2$ is a constant energy shift enforcing continuity at $\v P=\v P_{\rm crit}$. The effective mass $m^{*}$ of the polaron depends on the strength of the interaction between impurity and Bose atoms while $c$ is the speed of sound in the BEC which depends on the strength of the interaction between Bose atoms. We see that on the polaron side of this transition ($\abs{\v P}<\abs{\v P_{\rm crit}}$), the FMGS energy is quadratic with total momentum, with the polaron's effective mass $m^{*}$ being defined as the curvature of the energy around $\abs{\v P}=0$. On the Cherenkov side of this transition ($\abs{\v P}>\abs{\v P_{\rm crit}}$), the energy depends linearly on total momentum. 
The slope of the linear spectrum gives us the `polaron velocity', which is the semiclassical velocity that a polaron FMGS needs to reach in order to enter the Cherenkov regime. We find this velocity equals the average velocity of the impurity as it must, and both are equal to the speed of sound $c$ regardless of interaction strength. Momentum in this regime is shed into Bogoliubov excitations of the BEC. The dispersion relation in Eq.~\eqref{eq:PolDispersion} should hold even for large total momenta as long as the impurity-boson interaction can be modeled using a contact interaction as discussed in Sec.~\ref{sec:System}.

By examining the full distribution function for individual phonons and the impurity's momentum (describing quantum fluctuations around the mean), we see evidence of quasiparticle breakdown at the transition, indicated by a vanishing quasiparticle residue. The phonons present in this regime have very low momentum and there is a dramatic increase in the number of phonons at the transition; the system redistributes momentum into these phonons in order to fix the impurity's average velocity to the BEC speed of sound. It is the large number of phonons in the Cherenkov regime that makes the quasiparticle residue vanish.

We identify the coherent part of the impurity's momentum distribution, represented by a delta function peak at the value of the total system momentum. The weight contained in this part corresponds to the quasiparticle residue in Eq. \eqref{Obs_gs-Z}. The shape of the remaining incoherent part of the distribution is illustrated in Fig. \ref{fig:GS_PhaseDiagram}. These features of the impurity's momentum distribution can be directly probed in time-of-flight measurements.

We find that the quasiparticle residue sharply drops to zero at the critical momentum, which we take as a signature of quasiparticle breakdown in the supersonic regime. We also see the incoherent part of the distribution become sharply peaked at the critical momentum before broadening out again. This can be understood by the reasoning given in \cite{Nielsen2019} that at the critical momentum, the impurity travels at the same speed as the phonon excitations and therefore the phonons cannot dissipate momentum away. Another way to think about this effect is that the available phase space to excite phonons vanishes as the impurity approaches the speed of sound \cite{Boyanovsky2019}.

We establish that the FMGS as a function of total momentum has a transition at a critical momentum set by the effective mass of the polaron in the subsonic regime (see Fig. \ref{fig:GS_PhaseDiagram}). The transition occurs when the polaron velocity reaches the speed of sound. 

(II) \textit{Dynamical Transition}. We can also run the equations of motion in real time for various subsonic and supersonic momenta and examine the behavior of the system. The Cherenkov transition of the FMGS manifests in the dynamical behavior of various observables here. The Loschmidt Echo (dynamical equivalent of the quasiparticle residue) remains finite in the subsonic regime but has a power law decay to zero at long times in the Cherenkov regime for weak and intermediate interactions. This behavior is due to a logarithmic increase of total emitted phonon number at long times in the supersonic regime. Nielsen et. al and Boyanovsky et. al also predict this behavior in the Cherenkov regime \cite{Nielsen2019, Boyanovsky2019}, but find an exponential decay instead of a power law; we attribute this difference to their use of the Markov approximation.

The dynamical transition is also evident in the behavior of the average impurity velocity. Initially supersonic impurities have an average impurity velocity which exhibit a power-law decay towards $m^{*} c$ at long times, though convergence is slow; this is related to the divergence in the damping rate described in \cite{Nielsen2019, Boyanovsky2019} as the impurity approaches the critical momentum and cannot dissipate momentum through the phonon excitations. We therefore see that the Loschmidt echo and average impurity velocity both exhibit power-law decays in the supersonic regime with the onset of this behavior depending on interaction strength through a polaronic mass renormalization effect.



\begin{figure*}[t!]
	\centering
	\includegraphics[width=0.99\textwidth]{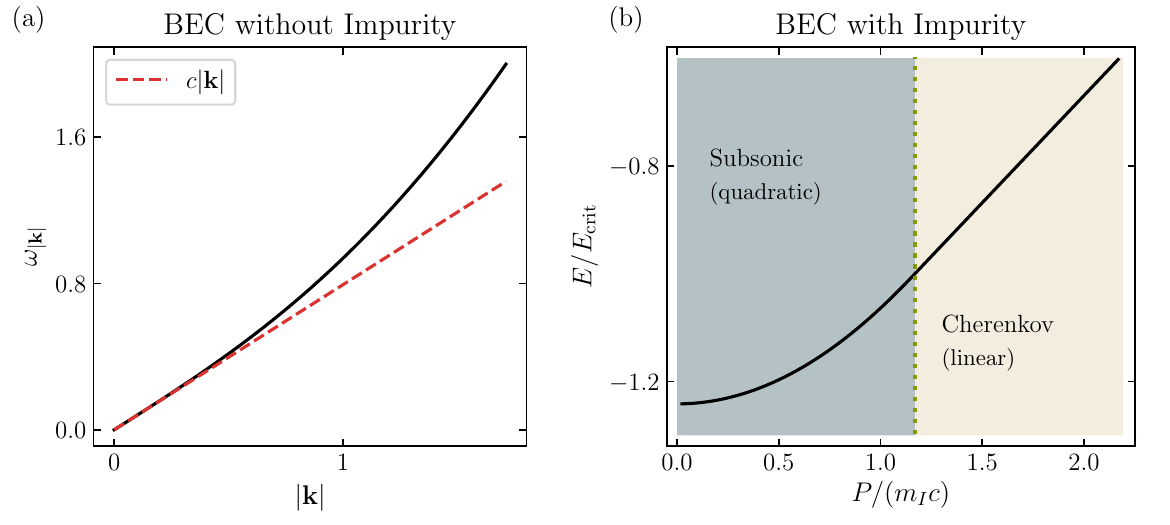}
	\caption{(a) The spectrum of Bogoliubov excitations (`phonons') in the BEC; the low energy spectrum is linear with such phonons traveling at the BEC speed of sound $c$. At larger momenta, the phonon dispersion becomes quadratic. Note that units are chosen such that $n_0=1$, $m_B=1$, and $\hbar=1$. (b) FMGS energy of the interacting impurity-BEC system in units of the energy at the critical momentum ($E_\textrm{crit}=E\left(\vec{P}_\textrm{crit}\right)$). We find that the spectrum is initially quadratic (subsonic regime) and then becomes linear (Cherenkov regime) at a critical momentum (green dotted line), after which momentum is shed into Bogoliubov excitations. The FMGS energy-momentum relation therefore behaves in an opposite manner to the phonon dispersion which is first linear and then smoothly becomes quadratic. For panel (b), the FMGS energy is computed for $a_{\rm IB}^{-1}=-4.46/\xi$ and a mass ratio of $m_{I}/m_{B}=1$. We use a momentum space UV cutoff $\Lambda=9.27/\xi$.}
	\label{fig:SchematicDispFig}
\end{figure*}

\section{System}
\label{sec:System}

In this section we define our system. We study an impurity immersed into a weakly interacting BEC of ultracold atoms near the inter-species Feshbach resonance. The system is in a three-dimensional box, $d=3$, of length $L$ in each dimension, and periodic boundary conditions are imposed. In our numerics, we find that $L$ in the range from $20\xi$-$70\xi$, where $\xi$ is the healing length given in Eq.~\eqref{eq:healingLength}, is sufficiently large to study observables without finite size effects.  

\subsection{Hamiltonian}

The microscopic Hamiltonian of entire system,
\begin{equation}\label{eq:H0}
\hat{H} = \hat{H}_{0} + \hat{H}_{\rm IB}
\end{equation}
consists of two parts: 
\begin{multline}\label{eq:H0_details}
\hat{H}_{0} =\sum_{\vec{k}}\varepsilon_{\vec{k}}^{I}\hat{d}_{\vec{k}}^{\dagger}\hat{d}_{\vec{k}} + \sum_{\vec{k}}\varepsilon_{\vec{k}}^{B}\hat{a}_{\vec{k}}^{\dagger}\hat{a}_{\vec{k}}\\ +\frac{1}{L^{d}}\frac{g_{\rm BB}}{2}\sum_{\vec{k},\vec{k'},\vec{q}}\hat{a}_{\vec{k}+\vec{q}}^{\dagger}\hat{a}_{\vec{k'}-\vec{q}}^{\dagger}\hat{a}_{\vec{k'}}\hat{a}_{\vec{k}}
\end{multline}
represents free impurities in an interacting Bose gas, and
\begin{equation}\label{eq:H_IB}
\hat{H}_\mathrm{IB}  =\frac{1}{L^d}g_{\rm IB}\sum_{\vec{k},\vec{k'},\vec{q}}\hat{d}_{\vec{k'}+\vec{q}}^{\dagger}\hat{d}_{\vec{k}+\vec{q}}\hat{a}_{\vec{k}}^{\dagger}\hat{a}_{\vec{k'}}
\end{equation}
represents the interaction between the impurities and the Bose gas. The boson creation (annihilation) operator in the momentum space is $\hat{a}^\dagger_{\vec{k}}$ ($\hat{a}_{\vec{k}}$); the same for the impurity is  $\hat{d}^\dagger_{\vec{k}}$ ($\hat{d}_{\vec{k}}$). The boson, $\varepsilon_{\vec{k}}^{B}$, and impurity, $\varepsilon_{\vec{k}}^{I}$, kinetic energies, are
\begin{equation}
\varepsilon_{\vec{k}}^{B}=\frac{\vec{k}^2}{2m_{B}}, \qquad \varepsilon_{\vec{k}}^{I}=\frac{\vec{k}^2}{2m_{I}}.
\end{equation}
Here, $m_B$ ($m_I$) stands for the boson (impurity) mass. The microscopic Hamiltonian, as defined by Eq.~\eqref{eq:H0} contains an arbitrary number of the impurity particles. For our work, however, we are interested in the states containing only one impurity. 

The Hamiltonian~\eqref{eq:H0_details} can be transformed in according to the standard Bogoliubov theory~\cite{pitaevskii_book_BEC}; using the linear transformation
\begin{equation}\label{eq:bogoMode}
\hat{a}_{\vec{k}} =u_{\vec{k}}\hat{b}_{\vec{k}} -v_{\vec{k}} \hat{b}_{\vec{-k}}^{\dagger}, \qquad \vec{k}\ne 0
\end{equation}
with the coefficients $u_\vec{k},v_{-\vec{k}} = \sqrt{\frac{\varepsilon_{\vec{k}}^{B} +g_\mathrm{BB} n_0}{2\omega_{\vec{k}}} \pm\frac12}$ we get
\begin{equation}
\hat{H}_{0} =\sum_{\vec{k}}\varepsilon_{\vec{k}}^{I}\hat{d}_{\vec{k}}^{\dagger}\hat{d}_{\vec{k}} + \sum_{\vec{k}}\omega_\vec{k} \hat{b}_{\vec{k}}^{\dagger}\hat{b}_{\vec{k}}
\end{equation}
for the terms in Eq.~\eqref{eq:H0_details} relative to the FMGS energy of the condensate. The operators $\hat{b}^\dagger_\vec{k}$ ($\hat{b}_\vec{k}$) create (annihilate) Bogoliubov quasiparticles with momentum $\vec{k}$ and dispersion
\begin{equation} \label{eq:omegaBEC}
\omega_{\vec{k}}=\sqrt{\varepsilon_{\vec{k}}^{B}(\varepsilon_{\vec{k}}^{B}+2g_\mathrm{BB}n_{0})}
\end{equation}
as depicted in Fig.~\ref{fig:SchematicDispFig}(a). Here, $n_{0}=N/L^{d}$ is the gas density (we let it be equal to the condensate density for the observables discussed in the paper), and $N$ is the number of bosons. The dispersion~\eqref{eq:omegaBEC} yields the sound velocity
\begin{equation}\label{eq:cBEC}
c=\sqrt{\frac{g_\mathrm{BB}n_{0}}{m_{B}}}.
\end{equation}
It also sets the healing length
\begin{equation}\label{eq:healingLength}
    \xi = (2m_{B} g_\mathrm{BB} n_{0})^{-1/2}
\end{equation}
at which the transition between the particle and phonon regimes of the BEC excitations occurs. The $s$-wave scattering length, $a_\mathrm{BB}\ge 0$, sets the value of the coupling strength $g_\mathrm{BB}$ in Eq.~\eqref{eq:H0_details}:
\begin{equation}\label{eq:gBB}
g_\mathrm{BB} =  \frac{4\pi a_\mathrm{BB}}{m_{B}}.
\end{equation}

The Hamiltonian~\eqref{eq:H_IB} represents the impurity-boson interaction potential in a way that is uniform in momentum space and, therefore, requires a regularization to be physically meaningful. We implement the regularization procedure commonly used in the existing literature~\cite{Rath2013, Shashi2014, Shchadilova2016}: we impose a sharp ultraviolet (UV) cutoff $\Lambda$ onto the the momentum space of the problem. The interaction parameter $g_\mathrm{IB}$ and the impurity-boson scattering length $a_\mathrm{IB}$ are connected through the Lippmann-Schwinger equation
\begin{equation}\label{eq:LippSchwing}
g_{\rm IB}^{-1} =  \frac{\mu}{2\pi} a_{\rm IB}^{-1} -\frac{1}{L^d} \sum_{\v k}^\Lambda \frac{2\mu}{\v k^2}.
\end{equation}
Here, 
\begin{equation}
	\mu=\frac{m_I m_B}{m_I+m_B}
\end{equation}
is the reduced mass of the two-body impurity-boson system. We stress that the second term in Eq.~\eqref{eq:LippSchwing} renormalizes $g_\mathrm{IB}$, thus giving a meaning to the interaction term~\eqref{eq:H_IB} in the $\Lambda\to\infty$ limit. The observables from our paper are evaluated for $\Lambda$ large enough to be viewed as if we took the $\Lambda\to\infty$ limit.

\subsection{Lee-Low-Pines transformation}

Now we take our model defined in the laboratory reference frame and write it in the mobile impurity reference frame. These reference frames are connected with the Lee-Low-Pines transformation,
\begin{equation}
\hat{\mathcal{S}}= e^{i\hat{\vec{R}}_\textrm{imp}\hat{\vec{P}}_\textrm{ph}},
\end{equation}
named after the original work~\cite{Lee1953}. Here, $\hat{\vec{R}}_\textrm{imp}$ is the position operator of the impurity. Any operator $\hat{\mathcal{O}}$ defined in the laboratory frame takes the form
\begin{equation}
\hat{\mathcal{O}}_\textrm{LLP} = \hat{\mathcal{S}} \hat{\mathcal{O}} \hat{\mathcal{S}}^\dagger
\end{equation}
in the mobile impurity reference frame. Note that we constrain our system to the case of a single impurity.

The total momentum of the system
\begin{align}\label{eq:totMom}
\hat{\vec{P}} & = \hat{\vec{P}}_{\rm imp} + \hat{\vec{P}}_{\rm ph}
\end{align}
is conserved.  In the above expression, the impurity momentum $\hat{\vec{P}}_{\rm imp}$ and total phonon momentum $\hat{\vec{P}}_{\rm ph}$ are given as
\begin{align}\label{eq:impPhonMom}
\hat{\vec{P}}_{\rm imp} & = \sum_{\v k} \v k \hat d^\dag_{\v k}\hat d_{\v k}\\
\hat{\vec{P}}_{\rm ph} & = \sum_{\v k} \v k \hat b^\dag_{\v k}\hat b_{\v k}.\label{eq:avgPhononMom}
\end{align}

The final Hamiltonian $\hat{H}_{\rm LLP}$ after the transformation reads:
\begin{equation}\label{eq:HLLP}
\hat{H}_{\rm LLP} \equiv \hat{\mathcal{S}} \hat{H}\hat{\mathcal{S}}^{\dagger}
=\hat{H}_{0, \rm LLP}+\hat{H}_{\rm IB, \rm LLP}
\end{equation}
where
\begin{equation}\label{eq:H_0LLP}
\hat{H}_{0, \rm LLP} =\sum_{\vec{k}}\omega_{\vec{k}}\hat{b}_{\vec{k}}^{\dagger}\hat{b}_{\vec{k}} + \frac{1}{2m_{I}}\left(\hat{\vec{P}}_{\rm imp}-\sum_{\vec{k}}\vec{k}\hat{b}_{\vec{k}}^{\dagger}\hat{b}_{\vec{k}}\right)^{2}
\end{equation}
and
\begin{multline} \label{eq:H_IBLLP}
\hat{H}_{\rm IB, \rm LLP}  =g_{\rm IB}n_{0} + g_{\rm IB}\sqrt{n_{0}}\frac{1}{\sqrt{L^{d}}}\sum_{\vec{k}\neq0}W_{\vec{k}}\left( \hat{b}_{\vec{k}}+\hat{b}_{-\vec{k}}^{\dagger}\right) \\
 +g_{\rm IB}\frac{1}{L^{d}}\sum_{\vec{k}\neq0,\vec{k'}\neq0}V_{\vec{k},\vec{k'}}^{\left(1\right)}\hat{b}_{\vec{k}}^{\dagger}\hat{b}_{\vec{k}'} \\
 +\frac{1}{2}g_{\rm IB}\frac{1}{L^{d}}\sum_{\vec{k}\neq0,\vec{k'}\neq0}V_{\vec{k},\vec{k}'}^{\left(2\right)}\left(\hat{b}_{-\vec{k}}\hat{b}_{\vec{k'}}+\hat{b}_{\vec{k}}^{\dagger}\hat{b}_{-\vec{k}'}^{\dagger}\right)
\end{multline}
Here, the interaction vertices satisfy the relations 
\begin{align}
W_{\vec{k}}&= \sqrt{\frac{\varepsilon_{\vec{k}}}{\omega_{\vec{k}}}}\\
V_{\v k\v k'}^{(1)} \pm V^{(2)}_{\v k \v k'} &= \left( W_{\v k} W_{\v k'}\right)^{\pm 1}
\end{align}
Note that the lab frame impurity momentum operator $\hat{\vec{P}}_{\rm imp}$ that appears in Eq.~\eqref{eq:H_0LLP} commutes with the Hamiltonian $\hat{H}_{\rm LLP}$ in the Lee-Low-Pines frame. We recognize it as the total momentum $\hat{\vec{P}}$ that commutes with the lab frame Hamiltonian $\hat{H}$ given in Eq. \eqref{eq:H0}. Therefore, we can replace $\hat{\vec{P}}_{\rm imp}$ with the quantum number $\vec{P}$; dynamics under this model will decouple into sectors indexed by $\vec{P}$. As the system is spherically symmetric, there is no preferred orientation of $\vec{P}$ and hence physical quantities will only depend on the magnitude
\begin{equation}\label{eq:Pmag}
    P=\abs{\vec{P}}.
\end{equation}
The first term in Eq. \eqref{eq:H_0LLP} describes a weakly interacting BEC where excitations are Bogoliubov phonons with dispersion $\omega_{\vec{k}}$. Enacting the Lee-Low-Pines transformation allows us to eliminate impurity degrees of freedom in the Hamiltonian at the cost of introducing an additional interaction term between phonons that show up in second term of Eq. \eqref{eq:H_0LLP}. The effective interaction strength $\frac{1}{2m_{I}}$ of this term is set by the impurity mass and vanishes for heavy impurities as the co-moving frame of the impurity and the original lab frame coincide.
 
 After the Lee-Low-Pines transformation, the part of the original Hamiltonian representing the impurity-boson interaction turns into Eq. \eqref{eq:H_IBLLP} which contains terms describing impurity-mediated interactions between the bosons. The linearized part of Eq. \eqref{eq:H_IBLLP} constitutes the Fr\"ohlich model which is suitable for describing weakly interacting polaronic systems. The term linear in phonon operators describes the impurity exciting phonons directly out of the condensate. If the impurity interacts strongly with the bosonic bath, it can also mediate scattering between phonon excitations; this is the physics included in the quadratic terms in Eq. \eqref{eq:H_IBLLP} and these terms are required for the study of strongly interacting polaronic systems beyond the Fr\"ohlich model \cite{Shashi2014}.

\subsection{Equations of motion}
\label{subsec:EoM}
We derive equations of motion for the system for the system state by approximating the true wavefunction with a single variational wavefunction and then time-evolving this variational state according the the Hamiltonian~\eqref{eq:HLLP}. Formally, we use Dirac's time-dependent variational principle to derive equations of motion for the coefficients defining the variational state ~\cite{Jackiw1979}. The FMGS is found by evolving the variational state in imaginary time while quench dynamics can be studied by evolving the state in real time. Our trial wavefunctions are coherent states of the form 
\begin{equation} \label{eq:WF}
\ket{\Psi_\textrm{coh}(t)} = e^{\sum_{\v k} \beta_{\v k}(t) \hat{b}_{\v k}^\dag - \rm H.c. } \ket{0}
\end{equation}
as used in the literature (e.g. ~\cite{Shashi2014, Shchadilova2016, VanLoon2018}). Note that $\rm H.c.$ means Hermitian conjugate. Here, $\beta_{\v k }(t)$ are the coherent state amplitudes and $\ket{0}$ denotes the  vacuum of Bogoliubov phonons (which is the FMGS of the BEC) in the co-moving frame of the impurity. We want to derive equations of motion for the coherent state amplitudes corresponding to time-evolution in the Lee-Low-Pines frame
\begin{equation}\label{eq:psit_LLP}
    \ket{\Psi_\textrm{coh}(t)} = e^{-i \hat{H}_{\rm LLP} t}\ket{0}.
\end{equation}
To derive the equations of motion, we first construct the classical action $\mathcal{S}=\int \mathcal{L}(t)dt$ where
\begin{align}
\mathcal{L}(t)=\bra{\Psi_\textrm{coh}(t)}i\partial_t-\hat{H}_{\rm LLP}\ket{\Psi_\textrm{coh}(t)}
\end{align}
is a classical Lagrangian calculated by projecting the true many-body wavefunction onto the submanifold of Hilbert space spanned by our chosen class of trial wavefunctions. The least action principle then gives the standard Euler-Lagrange equations of motion
\begin{eqnarray}\label{EL}
\frac{d}{dt}\frac{\partial \mathcal L}{\partial \dot \beta_{\vec{k}}}-\frac{\partial \mathcal L}{\partial \beta_{\vec{k}}} &=& 0
\end{eqnarray}
which describe the dynamics of the variational parameters $\beta_{\vec{k}}(t)$ and $\phi(t)$. The explicit expressions for these equations of motion are
\begin{multline}\label{eq:Dyn}
i \dot \beta_{\v k} = g_{\rm IB} \sqrt{n_{0}} W_{\v k} +\Omega_{\vec{k}} \beta_{\v k} \\+ g_{\rm IB}\left(W_{\v k}\chi^{+}_{\beta}+W_{\v k}^{-1}\chi^{-}_{\beta}\right)
\end{multline}
where we have defined
\begin{align}
\chi^{\pm}_{\beta}&=\frac{1}{2}\sum_{\v k'} W_{\v k'}^{\pm 1}\left(  \beta_{\v k'}\pm  \beta_{\v k'}^*\right)\\
\Omega_{\vec{k}}&=\omega_{\v k} + \frac{\v k^2}{2m_I} - \frac{1}{m_I}\v k \left( {\v P} - \v P_{\rm{ph}}\right)
\end{align}
and $\v P_{\rm{ph}} = \sum_{\v k'} \v k' \abs{\beta_{\v k'}}^2$ follows from Eq. \eqref{eq:avgPhononMom} and Eq. \eqref{eq:WF}. 
 Similarly, we can relate the interaction parameter $g_{\rm BB}$ to the boson-boson scattering length $a_{\rm BB}$ by solving the two-body boson-boson scattering problem. As the boson-boson interaction is weak, we can use the Born approximation rather than the full Lippman-Schwinger equation to get Eq.~\eqref{eq:gBB}.

At this point, we are ready to study the FMGS or quench dynamics of the system by evolving the equations of motion in imaginary time or real time respectively. The FMGS can alternatively be studied by calculating the analytical saddle point solution to Eq. \eqref{eq:Dyn} as done in \cite{Shchadilova2016}. However, this approach does not allow us to study the Cherenkov regime as the saddle point solution assumes that the variational parameters $\beta_{\v k }$ are real; for large enough total momentum, this is not true.

\subsection{Observables}
\label{subsec:ObsDist}
The trial wavefunction $\ket{\Psi_\textrm{coh}(t)}$ is characterized by the coherent state amplitudes $\beta_{\v k}$; the square $\abs{\beta_{\vec{k}}}^{2}$ of these amplitudes can be used to calculate a number of physical quantities that we can use to understand the system. Recalling that $\vec{P}$ is the total conserved momentum of the system, two such quantities that are relevant to both the FMGS and quench dynamics are
\begin{equation}\label{Obs-Nph}
N_{\rm{ph}}(t) = \sum_{\vec{k}}\abs{\beta_{\vec{k}}}^{2}
\end{equation}
and
\begin{equation}
\v P_{\rm{ph}} = \sum_{\v k} \v k \abs{\beta_{\v k}}^2
\end{equation}
where expectations are taken over $\ket{\Psi_\textrm{coh}(t)}$. These quantities are the average total phonon number and the average total phonon momentum. There are also two distribution functions that are useful to understand the system:
\begin{align}
n_{\rm ph}(\vec{k}) &= \frac{1}{N_{\rm{ph}}}\left\langle \hat{b}_{\vec{k}}^{\dag}\hat{b}_{\vec{k}}  \right\rangle \label{eq:IndPhonMomDist}\\
&= \frac{1}{N_{\rm{ph}}}\abs{\beta_{\vec{k}}}^{2}\nonumber
\end{align}
and
\begin{align}
n_{\rm imp}(\vec{p}) &= \left\langle \delta\left(\hat{\vec{P}}_{\rm imp}-\vec{p}\right)  \right\rangle\label{eq:ImpMomDist}\\
& = e^{-N_{\rm{ph}}(t)}\delta\left(\vec{P} - \vec{p}\right) + \tilde{n}_{\rm imp}(\vec{p})\nonumber
\end{align}
where
\begin{equation}
\tilde{n}_{\rm imp}(\vec{p})=\frac{e^{-N_{\rm{ph}}(t)}}{(2\pi)^{3}}\sum_{\vec{r}}e^{-i(\vec{P}-\vec{p})\vec{r}}\left(e^{\sum_{\vec{k}}\abs{\beta_{\vec{k}}}^{2}e^{i\vec{k}\vec{r}}}-1\right). \label{eq:ImpMomDistIncoh} 
\end{equation}
The first function, Eq. \eqref{eq:IndPhonMomDist}, describes how individual phonon occupation is distributed over momentum space on average.
The second function, Eq. \eqref{eq:ImpMomDist}, describes the impurity's momentum distribution and contains two terms. The first term is a delta function at $\vec{p}=\vec{P}$ corresponding to a coherent part of the distribution while the second term, Eq. \eqref{eq:ImpMomDistIncoh}, represents the incoherent part of the distribution. Ther derivation of Eq. \eqref{eq:ImpMomDist} is given in Appendix \ref{appdx:DistDeriv}. We can straightforwardly compute the distribution for impurity momenta with magnitude $p\equiv\abs{\vec{p}}$ as
\begin{equation}
n_{\rm imp}(p)=\sum_{\vec{p'}}\delta\left(\abs{\vec{p'}}-p\right)n_{\rm imp}(\vec{p'})
\end{equation}
which corresponds to examining spherical shells of the full three-dimensional distribution $n_{\rm imp}(\vec{p})$.

When examining the FMGS specifically, it is also useful to examine the energy and quasiparticle residue which are given in Eq.~\eqref{Obs_gs-E} and Eq.~\eqref{Obs_gs-Z} below:

\begin{multline}\label{Obs_gs-E}
E = \left\langle\hat{H}_{\rm LLP}\right\rangle
= \frac{1}{2m_I}\left(\vec{P}^2 -\v P_{\rm{ph}}^2 \right) + \sum_{\vec{k}}\Omega_{\vec{k}} \abs{\beta_{\v k}}^2\\
+ g_{\rm IB}\left(\chi^{+}_{\beta}+\sqrt{n_{0}}\right)^2-g_{\rm IB}\left(\chi^{-}_{\beta}\right)^2
\end{multline}
\begin{equation}\label{Obs_gs-Z}
Z = \abs{\braket{0|\Psi_\textrm{coh}^{\rm gs}}}^{2}=e^{-N_{\rm{ph}}}.
\end{equation}
where the average is taken over the FMGS trial wavefunction $\ket{\Psi_\textrm{coh}^{\rm gs}}$ and $\ket{0}$ is the state corresponding to a non-interacting impurity immersed in a BEC. The quasiparticle residue characterizes how much bare impurity character is leftover in the interacting FMGS; it can therefore be used to probe the breakdown of the quasiparticle picture. A natural generalization of the FMGS quasiparticle residue to the case of quench dynamics is the Loschmidt echo
\begin{equation}\label{eq:Obs-LS}
S(t) = \braket{0| e^{i\hat{H}_{\rm free}t}e^{-i\hat{H}_{\rm LLP}t}|0}
\end{equation}
where
\begin{equation}
\hat{H}_{\rm free}=\sum_{\vec{k}}\omega_{\vec{k}}\hat{b}_{\vec{k}}^{\dagger}\hat{b}_{\vec{k}} + \frac{\hat{\vec{P}}^{2}}{2m_{I}}
\end{equation}
is the Hamiltonian for the free impurity immersed in a BEC. Utilizing Eq.~\eqref{eq:psit_LLP}, we have
\begin{equation}\label{eq:Obs-LS2}
S(t) = e^{i\left[\frac{\vec{P}^2}{2m_{I}}t\right]}e^{-\frac{1}{2}N_{\rm{ph}}(t)}.
\end{equation}
The Loschmidt echo therefore represents the overlap between the state during time evolution and the original bare impurity. The Fourier transform of the echo can also be used to compute the RF absorption spectrum \cite{Shashi2014}.

The FMGS quasiparticle residue $Z$ is connected to the Loschmidt echo at infinite time
\begin{equation} \label{eq:SinfZ}
\left|S\left(t_{\infty}\right)\right| \equiv\lim_{t\to \infty}|S(t)| = Z.
\end{equation}
after an initial quench from a noninteracting state $\ket{0}$. The details of the derivation are given in Appendix~\ref{appdx:LEchoDeriv}.

\section{FMGS Transition}
\label{sec:GroundState}

\begin{figure*}[t!]
	\centering
	\includegraphics[width=0.97\textwidth]{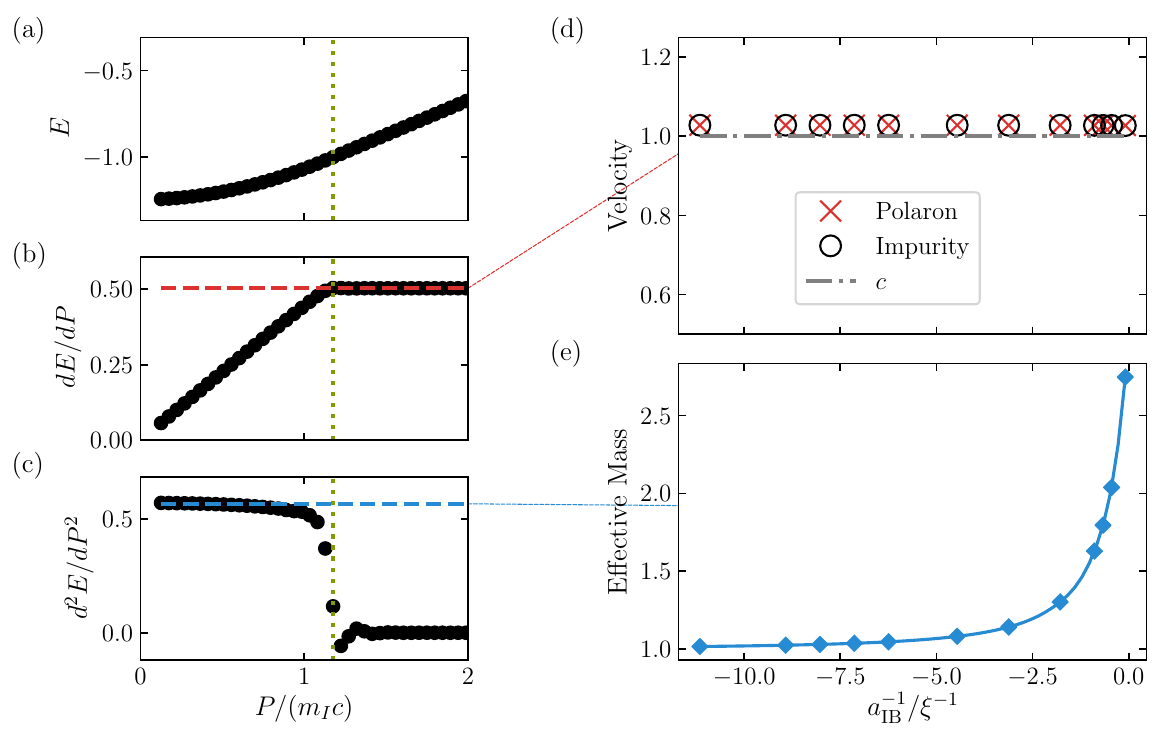}
	\caption{FMGS energy of an impurity immersed in a weakly interacting BEC. (a) The FMGS energy is initially quadratic (subsonic
		regime) and then becomes linear (Cherenkov regime) above a critical momentum where the critical momentum is indicated by the vertical green dotted line. The energy is plotted in units of $\abs{E_{\rm crit}}$ which is the value of energy at critical momentum. Panels (b) and (c) show the first and second derivatives of the energy respectively; they confirm the functional form of the system's energy-momentum relation hypothesized in Eq. \eqref{eq:PolDispersion}. These two curves are also normalized by $\abs{E_{\rm crit}}$. Note that panels (a), (b), and (c) correspond to an interaction strength $a_{\rm IB}^{-1}=4.46/\xi$. (d) Polaron group velocity and average
		velocity of the impurity in the Cherenkov regime (in units of $c$). The polaron velocity is given by the red `x's; each `x' corresponds to the constant plateau illustrated by the dashed red line in panel (b) which is computed for a specific interaction strength. The average impurity velocity is computed independently and confirmed to equal the polaron velocity as the Hellmann-Feyman theorem indicates. We see that the polaron velocity equals the BEC's speed of sound regardless of interaction strength, thus confirming half of the energy-momentum relation we hypothesis in Eq. \eqref{eq:PolDispersion}. (e) Effective mass of polaron in units of $m_{I}$. Each point corresponds to the average value near $P\sim 0$ illustrated by the dashed blue line in panel (c) which is computed for a specific interaction strength. The effective mass increases with interaction
		strength, but the group velocity does not. The mass ratio is $m_{I}/m_{B}=1$ and we use a sharp UV cutoff $\Lambda=9.27[\xi^{-1}]$.}
	\label{fig:GS_EnergyDerivsVelMass}
\end{figure*}

Here, we perform evolution of the system in imaginary time $\tau$ by running  Eq.~\eqref{eq:Dyn} with $t \to \tau=-it$ in order to numerically determine the FMGS. We examine characteristics of the FMGS of the system as we vary the total system momentum $\vec{P}$. We find that the FMGS energy initially has quadratic dependence on $\vec P$ for small momenta and then undergoes a transition to linear dependence on $\vec P$. The critical momentum increases for larger impurity-boson interaction strengths. In Fig. \ref{fig:GS_PhaseDiagram}, we plot the phase diagram, with the dispersion of the ground state system depicted in Fig.~\ref{fig:SchematicDispFig}(b). Note that the system is spherically symmetric so physical quantities will only depend on the magnitude of the total momentum $P$, Eq.~\eqref{eq:Pmag}.

We posit that system's FMGS energy-momentum relation follows Eq. \eqref{eq:PolDispersion} with the critical momentum of the transition being given by Eq. \eqref{eq:Pcrit}. Examining the FMGS energy and its derivatives gives evidence for this hypothesis. In Fig. \ref{fig:GS_EnergyDerivsVelMass} (a)-(c), we confirm  that for a fixed interaction strength, the energy is clearly quadratic at low momenta and then becomes linear. The system therefore behaves as a massive quasiparticle with effective mass $m^{*}=\nabla_{\vec{P}}^{2}E$ before the critical momentum $\vec{P}_{\rm crit}$. We call this regime the subsonic regime. Fig \ref{fig:GS_EnergyDerivsVelMass} (e) shows the effective mass for different interaction strengths; we see that it increases for stronger interactions which is the expected mass-renormalized effect.

We call the linear energy-momentum regime the Cherenkov regime. In Fig. \ref{fig:GS_EnergyDerivsVelMass} (d), we plot the polaron's group velocity, defined as
\begin{equation}
\vec{v}_{\rm pol}\left(\vec{P}\right)=\nabla_{\vec{\tilde{P}}}E\biggr\rvert_{\vec{\tilde{P}}=\vec{P}}
\end{equation}
and find that it is equal to the BEC's sound speed~\eqref{eq:cBEC} regardless of interaction strength. The Hellmann-Feynman theorem tells us that the average impurity velocity defined as
\begin{equation}\label{eq:vimp_gs}
	\vec{v}_{\rm imp}\left(\vec{P}\right)=\frac{\braket{\hat{\vec{P}}_{\rm imp}}}{m_{I}}
\end{equation}
must be equal to the polaron's group velocity $\vec{v}_{\rm pol}$ which we also confirm in the figure. The simulation-derived polaron velocity and average impurity velocity are within $2\%$ of the BEC's sound speed. The low energy phonons that dress the impurity in the Cherenkov regime by definition have a group velocity $c$. We therefore have a physical picture in this regime of the impurity moving on average at the speed of sound along with low-energy
phonons traveling at the same speed. The fact that the group velocity of the low-energy part of the phonon dispersion equals the group velocity of the high-energy part of the total system is not obvious a priori.

Next, we examine the critical momentum at which the transition occurs. In Fig. \ref{fig:GS_PhaseDiagram}(a), we see that the transition happens at larger momentum for stronger interaction strengths. The transition line in the figure is computed from the discontinuity in the second derivative of the energy, corresponding to a continuous but not smooth dispersion. Note that the rapid but smooth change in the second derivative of the energy exhibited in our numerics is suggestive of a sharp transition but does not prove the existence of one. Increasing the size of the $L$ of the system (corresponding to the position space grid) is likely to shrink the finite window of the numerical transition. We also plot the predicted transition line given by Eq. \eqref{eq:Pcrit} and see that it reasonably matches the true transition. Thus, we conclude that the transition occurs when the polaron's group velocity reaches the BEC's sound speed; the amount of momentum needed for this to occur increases for stronger interactions due to the polaronic mass-renormalization of the impurity.

\begin{figure*}[t!]
	\centering
	\includegraphics[width=0.99\textwidth]{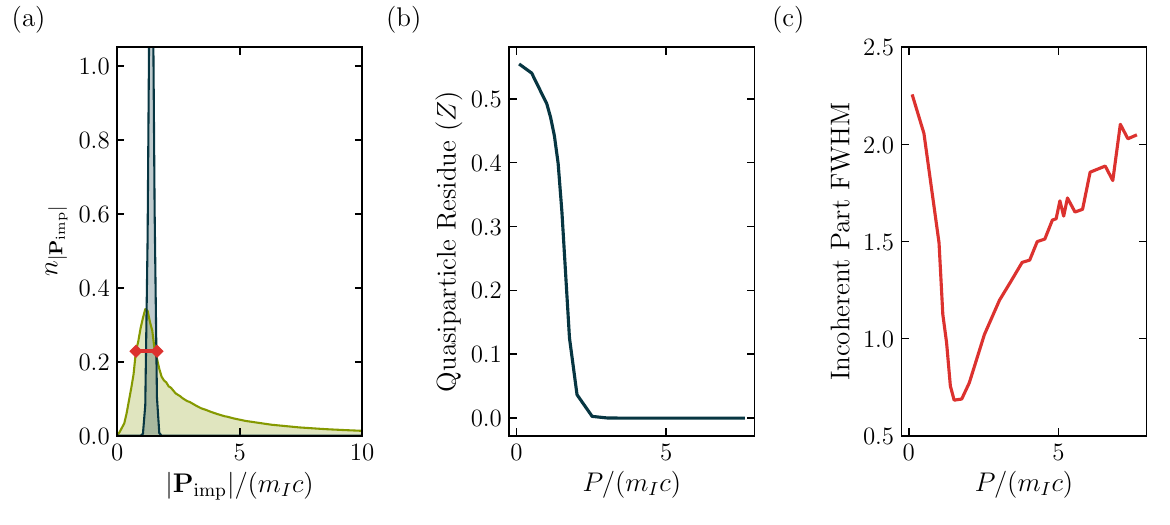}
	\caption{Impurity's speed distribution for $a_{\rm IB}^{-1}=-1.78/\xi$ and $P/(m_{I}c)=1.39$ (a) Example distribution showing the
		coherent $\delta$-peak of the distribution (dark blue line line) which has been Gaussian-broadened by hand and the incoherent part of the distribution (green). The red
		line shows the full-width half-max (FWHM) of the incoherent part.
		(b) Quasiparticle residue. The weight contained in the $\delta$-peak of the distribution is the quasiparticle residue which drops to zero at the transition. (c) FWHM
		of incoherent part. The sharp dip at the transition indicates that
		the incoherent part of the impurity's speed distribution gets sharply
		peaked at the critical momentum. The mass ratio is $m_{I}/m_{B}=1$ and we use a sharp UV cutoff $\Lambda=9.27/\xi$.}
	\label{fig:GS_nPImag}
\end{figure*}

We gain further insight into the transition by examining the impurity's momentum distribution. In Fig. \ref{fig:GS_nPImag}, we plot the weight of the coherent part of the distribution (which is equivalent to the polaron's quasiparticle weight) as well as the FWHM and height of the peak of the incoherent part of the momentum magnitude distribution. We see that the quasiparticle weight drops to zero at the transition, indicating quasiparticle breakdown. We can understand the
behavior of the quasiparticle residue as the system producing an `infinite'
number of low energy phonons in the Cherenkov regime and therefore
having a vanishing overlap with the bare non-interacting impurity. The incoherent part of the impurity's momentum distribution also gets sharply peaked at the transition before broadening out again. As stated in \cite{Nielsen2019}, the system almost acts like a momentum eigenstate of a non-interacting impurity at the transition as phonons cannot dissipate momentum away from the impurity; the available phase space of excitations vanishes as the impurity approaches the speed of sound \cite{Boyanovsky2019}.

In summary, the FMGS of the system exists either in a subsonic regime
where the system truly behaves as a dressed quasiparticle with an
effective mass, a quadratic energy-momentum relation, and non-zero quasiparticle
residue or in a Cherenkov regime characterized by the impurity traveling
at the speed of sound along with a large number of low energy phonons
and a vanishing quasiparticle residue (quasiparticle breakdown). The
critical momentum at which the transition occurs depends on the interaction
strength and corresponds to the polaronic mass renormalization of
how much momentum needs to be input into the system to get the impurity
to travel at the speed of sound. 

Physical intuition for the transition can be gained via an energy stability argument. If the system forms a quasiparticle with dispersion $\vec{P}^{2}/2m^{*}$, scattering with Bogoliubov excitation with momentum $\vec{k}$ and energy $\omega_{\vec{k}}$ is only energetically prohibited for $\vec{P}<m^{*}c$. A quasiparticle with linear dispersion $c\abs{\vec{P}}$ is stable to such excitations even for $\vec{P}>m^{*}c$ as long as the quasiparticle's velocity is not greater than $c$. A quasiparticle state with quadratic dispersion is therefore only stable for momenta below $\abs{\v P_{\rm crit}}=m^{*} c$ and undergoes a transition at this critical momentum. Below this critical point, the impurity can only be dressed by virtual excitations. At higher momenta, however, real excitations becomes accessible and thus dramatically change the nature of the dressed quasiparticle. The qualitative difference in dressing between the subsonic and Cherenkov regimes is evidenced by the quasiparticle residue dropping to zero in the latter.  

It is important to note that while the quasiparticle residue goes
strictly to zero in the Cherenkov regime regardless of interaction strength,
this effect becomes difficult to detect in practice at strong interactions where sufficiently large numbers of phonons are excited. Even in the subsonic regime where the residue is always finite, the large number of phonons makes this finite residue increasingly small. Therefore, at sufficiently strong
interactions, the quasiparticle residue by itself cannot be used to
distinguish between the subsonic regime and the Cherenkov regime.


\section{Dynamical Transition}
\label{sec:ReDyn}
We take a non-interacting impurity propagating through a BEC in its FMGS and instantly switch on (quench) the impurity-BEC interaction at time $t_{0}=0$. We then evolve the system by running Eq.~\eqref{eq:Dyn} in real time. Recall that the total momentum  $\vec{P}$ of the system is conserved and this momentum initially resides on the non-interacting impurity; the initial velocity of the impurity is
\begin{equation} \label{eq:v0}
\vec{v}_\mathrm{imp} (t_{0}) =\frac{\vec{P}}{m_{I}}.
\end{equation} 
Hereafter, the initial impurity velocity and the total system momentum are used interchangeably depending on the context of the discussion.

We find a qualitative change in the long-time value of the Loschmidt echo, Eq.~\eqref{eq:Obs-LS}, and of the average impurity velocity as $\vec{P}$ crosses a critical value $\vec{P}_\textrm{crit}^\textrm{dyn}$. We then discuss how $\vec{P}_{\textrm{crit}}^\textrm{dyn}$ is related to $\vec{P}_\textrm{crit}$ for the FMGS transition studied in section~\ref{sec:GroundState}.


\begin{figure*}[t!]
	\centering
	\includegraphics[width=\textwidth]{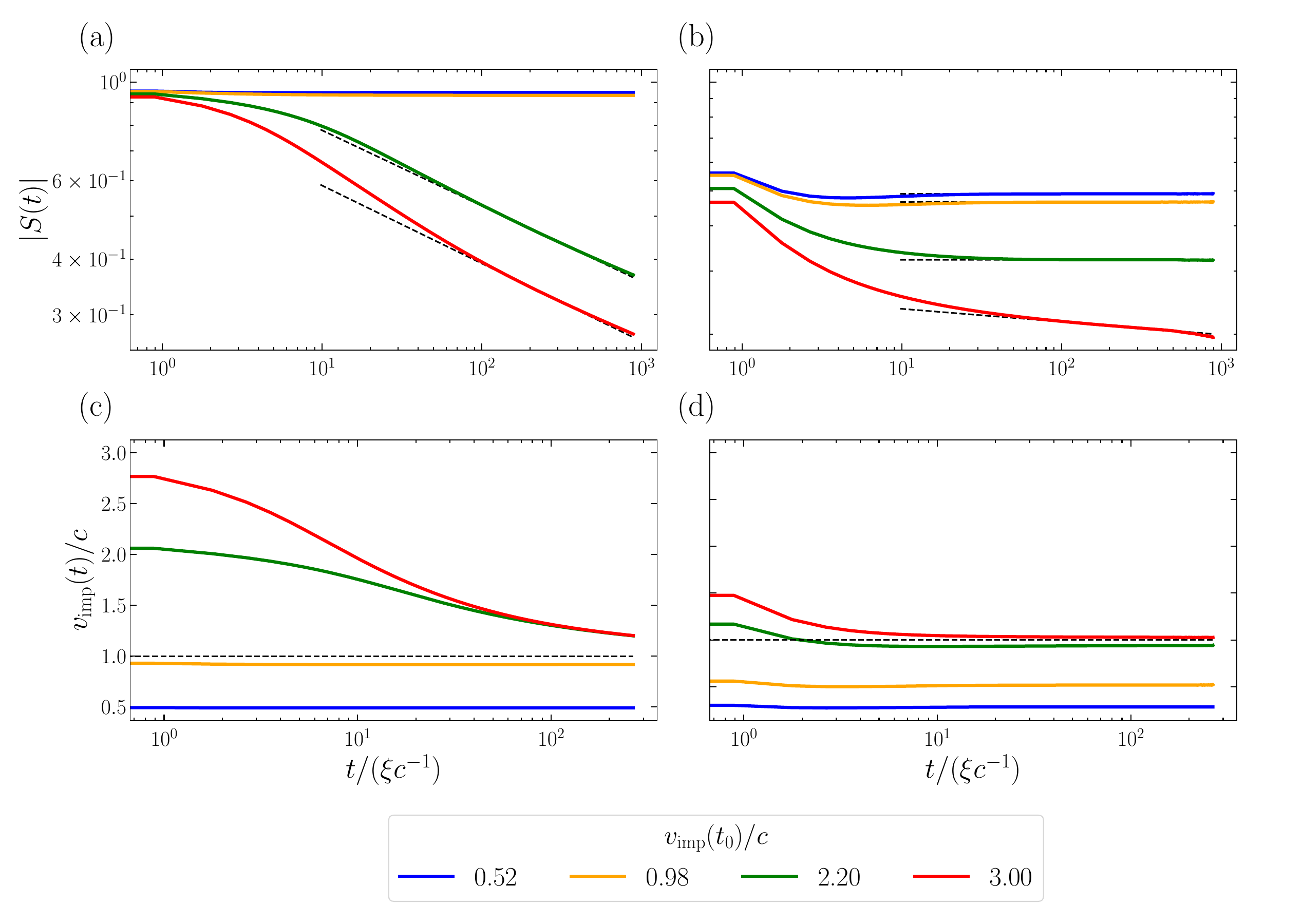}
	\caption{(a) Loschmidt echo at relatively weak interaction ($a_{\rm IB}^{-1}=-8.92/\xi$). Curves from the subsonic
		regime saturate to a finite value at long times. Curves in the Cherenkov
		regime decay to zero as a power-law at long times. (b) Loschmidt echo at strong interaction ($a_{\rm IB}^{-1}=-1.78/\xi$). Curves for sufficiently fast impurities also exhibit power-law decay. (c) Average speed of the impurity at relatively weak interaction ($a_{\rm IB}^{-1}=-8.92/\xi$). Curves from the subsonic regime saturate to different subsonic
		values at long times. Curves in the Cherenkov regime all approach
		the speed of sound at long times. (d) Average speed of the impurity at strong interaction ($a_{\rm IB}^{-1}=-1.78/\xi$). Curves approach the speed of sound for sufficiently fast impurity velocity. The mass ratio is $m_{I}/m_{B}=1$ and we use a sharp UV cutoff $\Lambda=13.91/\xi$.}
	\label{fig:RD_ObsTime}
\end{figure*}

\begin{figure}[t!]
	\centering
	\includegraphics[clip, width=0.99\columnwidth, height=0.7\textheight]{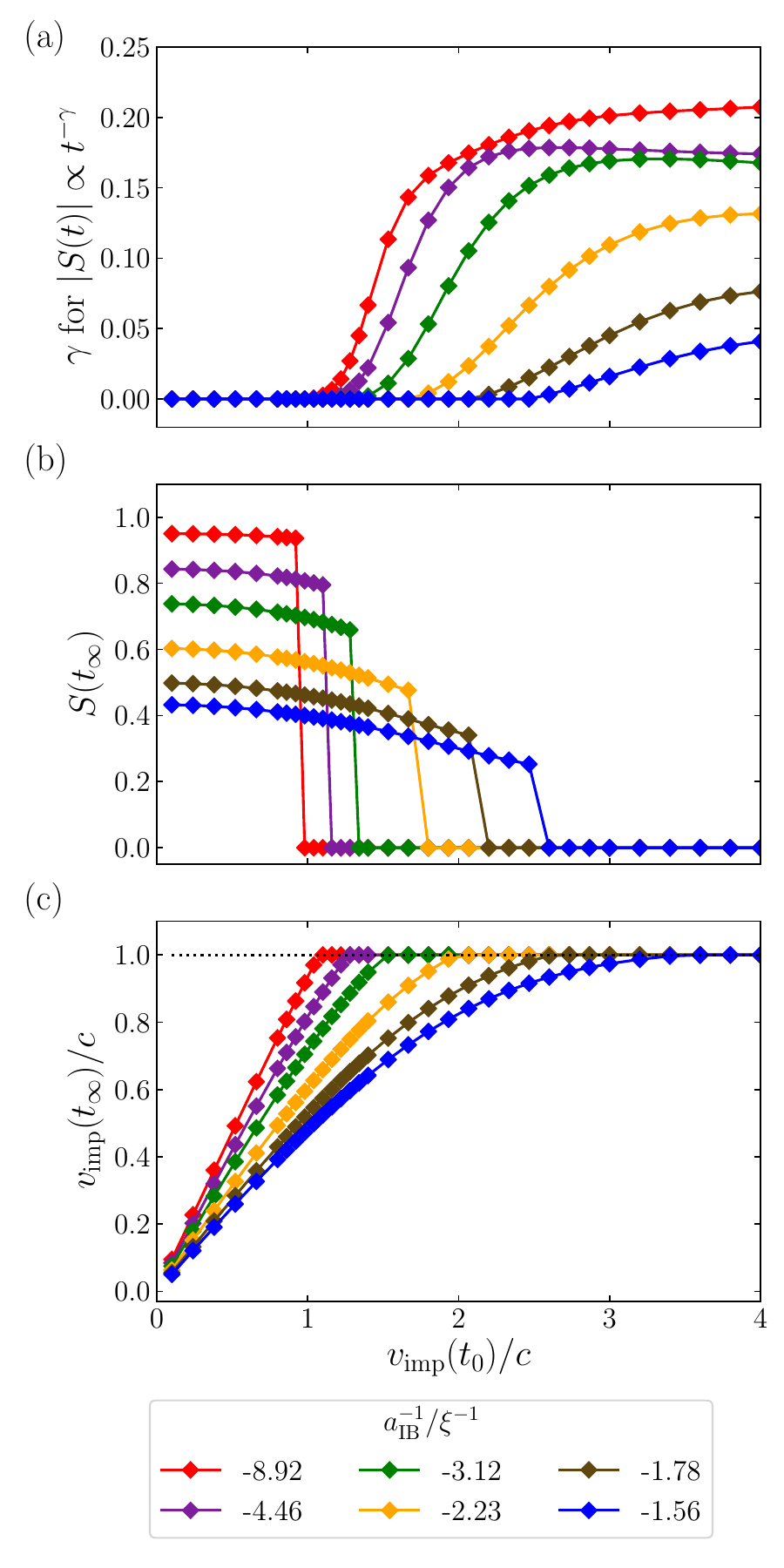}
	\caption{Asymptotic behavior of observables as a function of initial impurity velocity for mass ratio $m_{I}/m_{B}=1$. Note that the initial impurity velocity corresponds to the total momentum of the system which is conserved. (a) Exponents characterizing long-time behavior of the Loschmidt echo. The power-law decays only develop beyond a critical momentum which depends on interaction strength. (b) Long-time value of the Loschmidt echo. The long-time dynamical overlap is nonzero followed by a discontinuous drop to zero with this drop happening later for stronger interactions. (c) Long-time behavior of the average speed of the impurity. The impurity ends up at the speed of sound after a large enough initial velocity. We use a sharp UV cutoff $\Lambda=13.91/\xi$.}
	\label{fig:RD_ObsExpInfVals}
\end{figure}


\subsection{Loschmidt Echo}
\label{subsec:ReDyn_LSEcho}
We first examine the Loschmidt echo $S(t)$, Eq.~\eqref{eq:Obs-LS}. Figures~\ref{fig:RD_ObsTime}(a) and (b) illustrate its characteristic behavior for different initial impurity velocities, $\vec{v}_\mathrm{imp} (t_{0})$, at weak and strong interactions, respectively. We see that $|S(t)|$ saturates to a non-zero value at long times for small momenta and undergoes a power-law decay to zero as we cross some critical momentum $\vec{P}_\textrm{crit}^\textrm{dyn}$. This power-law decay in the dynamical overlap corresponds to a logarithmic growth in the phonon number in the supersonic regime: per Eq.~\eqref{eq:Obs-LS2}, we see that
\begin{equation}\label{eq:Nph_logDiv}
	N_\mathrm{ph}\left(t\right) \sim \left\{ \begin{array}{ll}
	\mathrm{constant}, & \text{if $\abs{\v P}\leq\abs{\vec{P}_\textrm{crit}^\textrm{dyn}}$}\\
	&\\
	\log{t}, & \text{if $\abs{\v P}>\abs{\vec{P}_\textrm{crit}^\textrm{dyn}}$}
	\end{array} \right.
\end{equation}
in the $t\to\infty$ limit. Physically, it is the divergence of $N_\mathrm{ph}$ which causes the decay of the Loschmidt echo. The increase in phonon number in the supersonic
regime is the essence of the Cherenkov radiation phenomenon. Its divergence with time is, however, not obvious \textit{a priori}. We expect that the logarithmic divergence in Eq.~\eqref{eq:Nph_logDiv} represents the leading order behavior of $N_\mathrm{ph}$, which is captured by the coherent state variational ansatz employed in our work. Therefore, going beyond our ansatz, for example, by taking into account squeezing correlations between phonon modes~\cite{Shi2017}, may only introduce subleading corrections to Eq.~\eqref{eq:Nph_logDiv}. Note that the works~\cite{Nielsen2019, Boyanovsky2019} also predict a decay of the Loschmidt echo in the supersonic regime, but find it be exponential instead of a power law. We attribute this difference to their use of the Markov approximation.


To find the value of the critical momentum $\abs{\vec{P}_\textrm{crit}^\textrm{dyn}}$, we fit the long-time tail of $|S(t)|$ with a power-law,
\begin{equation} \label{eq:Sfit}
|S(t)| = \mathrm{const} \times t^{-\gamma}, \qquad t\to\infty.
\end{equation}
We do the fit over the time range $80 \le t/\xi c^{-1} \le 90$ for all values of interactions available from our numerical simulation. The exponents $\gamma$ of these fits are shown in Fig.~\ref{fig:RD_ObsExpInfVals}(a) for various interaction strengths at a mass ratio $m_{I}/m_{B}=1$. Then, $\abs{\vec{P}_\textrm{crit}^\textrm{dyn}}$ is identified as the momentum upon which the exponents become nonzero. We see that the critical momentum always corresponds to an initial velocity of the impurity, Eq.~\eqref{eq:v0}, that is above the speed of sound,
\begin{equation} \label{eq:vimpcrit}
    \abs{\vec{v}_\mathrm{imp, crit}(t_{0})} = \frac{\abs{\vec{P}_\textrm{crit}^\textrm{dyn}}}{m_{I}} \ge c.
\end{equation}
The equality is only reached in the limit of vanishing impurity-boson interaction, and the value of $\abs{\vec{v}_\mathrm{imp, crit}(t_{0})}$ increases with the strength of the interaction.

Finally, we extrapolate our data for $|S(t)|$ to find the value of $|S(t_\infty)|$, Eq.~\eqref{eq:SinfZ}, for various initial impurity velocities by making use of the fitting formula~\eqref{eq:Sfit}. When the exponent $\gamma$, shown in Fig.~\ref{fig:RD_ObsExpInfVals}(a), is zero, $|S(t)|$ saturates to a finite value. When $\gamma$ is nonzero, $|S(t)|$ decays to zero. We illustrate this behavior in Fig.~\ref{fig:RD_ObsExpInfVals}(b). We see that $|S(t_\infty)|$ has a discontinuous drop to zero between adjacent data points as the initial impurity velocity crosses the value~\eqref{eq:vimpcrit}. 



The discontinuous behavior of $|S(t_\infty)|$ in the vicinity of $|\vec{P}^\textrm{dyn}_\textrm{crit}|$ identifies a dynamical transition. At the same time, $|S(t_\infty)|$ is determined by the FMGS quasiparticle residue $Z$, Eq.~\eqref{eq:SinfZ}. We find in section~\ref{sec:GroundState} that this residue has a discontinuity at $\abs{\vec{P}_\textrm{crit}}$ and therefore signals the subsonic-Cherenkov FMGS transition. A comparison between $|\vec{P}^\textrm{dyn}_\textrm{crit}|$ and $\abs{\vec{P}_\textrm{crit}}$ is given in Fig.~\ref{fig:GS_PhaseDiagram}(a), where we plot critical points of the dynamical transition on top of the FMGS phase diagram. We see that the qualitative behavior of the critical line is the same for this dynamical transition compared to the FMGS, but quantitatively there is disagreement at stronger interactions. Per Eq.~\eqref{eq:SinfZ}, however, we expect quantitative agreement of the transition line:
\begin{equation}\label{eq:DPTeqGST}
    |\vec{P}^\textrm{dyn}_\textrm{crit}|=\abs{\vec{P}_\textrm{crit}}.
\end{equation}
The fact that our numerical data does not exactly fulfil Eq.~\eqref{eq:DPTeqGST} at strong interactions may be due in part to different UV cutoffs being used for FMGS simulations versus quench dynamics; a smaller cutoff was used in the former to ensure that the imaginary time dynamics saturated within the limitations of available computational resources. Another contribution to the quantitative discrepancy may arise from limitations of the coherent state ansatz; higher order correlations between phonon modes are not explicitly captured and these correlations could become more relevant at stronger interactions. Using a full Gaussian state ansatz which includes squeezing on top of the coherent state could improve the agreement between the critical lines.


\subsection{Average Impurity Velocity}
\label{subsec:ReDyn_VImp}

In this section, we examine the behaviour of the average impurity velocity
\begin{equation}\label{eq:Obs-impVel}
	\vec{v}_\textrm{imp}\left(t\right)=\frac{\braket{\hat{\vec{P}}_\textrm{imp}(t)}}{m_{I}}.
\end{equation}
in an effort to further understand the dynamical transition. Recall that the FMGS transition is characterized independently by a discontinuity in the quasiparticle residue, Fig.~\ref{fig:GS_nPImag}(b), and a discontinuity in the second derivative of the energy, Fig.~\ref{fig:GS_EnergyDerivsVelMass}(c). Naturally, the average FMGS impurity velocity, Eq.~\eqref{eq:vimp_gs}, has a cusp at a point where the second derivative of the energy is discontinuous. This cusp can equivalently be used as a signature of the transition. The FMGS transition and dynamical transition are connected through the equivalence of the quasiparticle residue $Z$ and the asymptotic behavior of the Loschmidt echo $|S(t_\infty)|$, Eq.~\eqref{eq:SinfZ}. It is therefore natural to check whether there is a link between the FMGS impurity velocity and the asymptotic value 
\begin{equation}\label{eq:vimp_inf}
    \vec{v}_\textrm{imp}\left(t_{\infty}\right) \equiv\lim_{t\to \infty}\vec{v}_\textrm{imp}\left(t\right)
\end{equation}
of the average impurity velocity $\vec{v}_\textrm{imp}\left(t\right)$ after a quench. Unlike Eq.~\eqref{eq:SinfZ}, we do not have an analytic formula connecting $\vec{v}_\textrm{imp}\left(t_{\infty}\right)$ and the FMGS impurity velocity, Eq.~\eqref{eq:vimp_gs}. Our numerical analysis suggests that
\begin{equation}\label{eq:vimp_inf_asymp}
	\vec{v}_\textrm{imp}\left(t_{\infty}\right) = \left\{ \begin{array}{ll}
	\mathrm{constant}<c, & \text{if $\abs{\v P}\leq\abs{\vec{P}_\textrm{crit}^\textrm{dyn}}$}\\
	&\\
	c, & \text{if $\abs{\v P}>\abs{\vec{P}_\textrm{crit}^\textrm{dyn}}$}
	\end{array} \right.
\end{equation}
which indicates that the asymptotic average impurity velocity witnesses the dynamical transition, just as the ground stat impurity velocity witnesses the FMGS transition.

Our numerical analysis allowing the extrapolation of $\vec{v}_\textrm{imp}\left(t_{\infty}\right)$ is similar to the analysis of the Loschmidt echo in Sec.~\ref{sec:ReDyn}(a). Figures~\ref{fig:RD_ObsTime}(c) and (d) illustrate the characteristic behavior of $\vec{v}_\textrm{imp}\left(t\right)$ for different initial impurity velocities, $\vec{v}_\mathrm{imp} (t_{0})$, at weak and strong interactions, respectively. We see that the average impurity velocity decays to a subsonic value for smaller initial velocities, and decays towards the speed of sound for larger initial velocities. To find the asymptotic value of the impurity velocity, $\vec{v}_\textrm{imp}\left(t_{\infty}\right)$, we separately consider the case where the velocity, $\vec{v}_\textrm{imp}\left(t\right)$ with a power-law, has saturated to a value below $c$ in the simulated time window, and the case where it has not. In the former case, we take $\vec{v}_\textrm{imp}\left(t_{\infty}\right)$ to be the average of the last few values of $\vec{v}_\textrm{imp}\left(t\right)$. In the latter case, we fit the long-time tail of $\vec{v}_\textrm{imp}\left(t\right)$ with a power-law,
\begin{equation} \label{eq:vimpfit}
\vec{v}_\textrm{imp}\left(t\right) = c \times t^{-\gamma}, \qquad t\to\infty.
\end{equation}
We do the fit over the time range $80 \le t/(\xi c^{-1}) \le 90$ for all values of interactions available from our numerical simulation.

The extrapolated values $\vec{v}_\textrm{imp}\left(t_{\infty}\right)$ for various initial impurity velocities are shown in Fig.~\ref{fig:RD_ObsExpInfVals}(c). We find that the final impurity velocity is monotonic; it increases with initial impurity velocity until some critical $\abs{\vec{v}_\mathrm{imp, crit}(t_{0})}$ after which it saturates to $c$. For weak interactions, this critical initial velocity is seen to be the same as the critical transition point of the Loschmidt echo, cf. Fig~\ref{fig:RD_ObsExpInfVals}(b) and (c). For stronger interactions, however, there is a discrepancy in the critical initial velocities. We expect this discrepancy is due to the numerical difficulty of extrapolating $\vec{v}_\textrm{imp}\left(t_{\infty}\right)$.

\section{Conclusion}
\label{sec:Conclusion}

In this paper, we detail the behavior of a quantum Cherenkov transition in systems of finite momentum Bose polarons, with the transition existing for all impurity-boson interaction strengths and mass ratios~\cite{Seetharam_CherenkovPRL2021}. Experimentally, these systems are studied by immersing a dilute impurity gas in a BEC. We assume that the gases are trapped using a sufficiently large box trap, leading to a homogeneous BEC in the vicinity of the impurity.  The FMGS transition can be detected via the width of the impurity's momentum distribution extracted from time-of-flight imaging of the impurity gas. The dynamical transition should similarly be visible through time-of-flight imaging at late times. The onset of a power-law decay in the Loschmidt echo, measured via RF spectroscopy or Ramsey interferometery, can also be used as a signature of the dynamical transition. 

Ongoing  experiments  involving  harmonically-trapped dilute fermions immersed  in  an  oscillating BEC  are investigating the appearance of Cherenkov physics~\cite{Seetharam2022_thesis}. A proper theoretical analysis of such protocols, where the gases are harmonically trapped and subject to external forces, requires extending our formalism using the local-density approximation, which we present in a separate work~\cite{Seetharam_EffMassCherenkov2022}.

\section{Acknowledgments}
\label{sec:Acknowledgments}
We thank Carsten Robens, Martin Zwierlein, Zoe Yan, Yiqi Ni, Richard Schmidt, and Achim Rosch for helpful discussions. This research was conducted with Government support under and awarded by DoD, Air Force Office of Scientific Research, National Defense Science and Engineering Graduate (NDSEG) Fellowship, 32 CFR 168a. F. G. acknowledges support by the Deutsche Forschungsgemeinschaft (DFG, German Research Foundation) under Germany's Excellence Strategy -- EXC-2111 — 390814868. The work of M. B. Z. is supported by Grant No. ANR-
16-CE91-0009-01 and CNRS grant PICS06738. E. D. acknowledges support by the Harvard-MIT Center of Ultracold Atoms, ARO grant number W911NF-20-1-0163, and NSF EAGER-QAC-QCH award No. 2037687.


\appendix
\section{Distribution Function Derivation}
\label{appdx:DistDeriv}

The distribution of total phonon momentum is:
\begin{align}
n_{\rm ph,\text{tot}}(\vec{p}) &= \bra{\Psi_\textrm{coh}} \delta\left(\hat{\vec{P}}_{\rm{ph}}-\vec{p}\right)  \ket{\Psi_\textrm{coh}}\\
& =\braket{\psi|\frac{1}{\left(2\pi\right)^{3}}\sum_{\vec{r}}e^{i\vec{r}\left(\hat{\vec{P}}_{\rm{ph}}-\vec{p}\right)}|\psi}\\
& =\frac{1}{\left(2\pi\right)^{3}}\sum_{\vec{r}}e^{-i\vec{p}\cdot\vec{r}}e^{\sum_{\vec{k}}\left|\beta_{\vec{k}}\right|^{2}\left(e^{i\vec{k}\cdot\vec{r}}-1\right)}\\
& =\frac{1}{\left(2\pi\right)^{3}}\sum_{\vec{r}}e^{-i\vec{p}\cdot\vec{r}}e^{-N_{\rm ph}+\sum_{\vec{k}}\left|\beta_{\vec{k}}\right|^{2}e^{i\vec{k}\cdot\vec{r}}}
\end{align}
We can manipulate the expression into a more illuminating form as follows:
\begin{align}
n\left(\vec{p}\right) & =\frac{1}{\left(2\pi\right)^{3}}\sum_{\vec{r}} e^{-i\vec{p}\cdot\vec{r}}e^{-N_{\rm ph}+\sum_{\vec{k}}\left|\beta_{\vec{k}}\right|^{2}e^{i\vec{k}\cdot\vec{r}}}\\
& =\frac{e^{-N_{\rm ph}}}{\left(2\pi\right)^{3}}\sum_{\vec{r}} e^{-i\vec{p}\cdot\vec{r}}\left[e^{\sum_{\vec{k}}\left|\beta_{\vec{k}}\right|^{2}e^{i\vec{k}\cdot\vec{r}}}-1+1\right]\\
& =e^{-N_{\rm ph}}\frac{1}{\left(2\pi\right)^{3}}\sum_{\vec{r}} e^{-i\vec{p}\cdot\vec{r}}\\
& +\frac{e^{-N_{\rm ph}}}{\left(2\pi\right)^{3}}\sum_{\vec{r}} e^{-i\vec{p}\cdot\vec{r}}\left[e^{\sum_{\vec{k}}\left|\beta_{\vec{k}}\right|^{2}e^{i\vec{k}\cdot\vec{r}}}-1\right]\\
& =e^{-N_{\rm ph}}\delta\left(\vec{p}\right)+\tilde{n}\left(\vec{p}\right)
\end{align}
where the weight of the first term $e^{-N_{\rm ph}}$ corresponds to how connected the system is to the free impurity and the second term 
\begin{eqnarray}
\tilde{n}\left(\vec{p}\right)\equiv\frac{e^{-N_{\rm ph}}}{\left(2\pi\right)^{3}}\sum_{\vec{r}} e^{-i\vec{p}\cdot\vec{r}}\left[e^{\sum_{\vec{k}}\left|\beta_{\vec{k}}\right|^{2}e^{i\vec{k}\cdot\vec{r}}}-1\right]
\end{eqnarray}
corresponds to particle momenta distributed over an incoherent background.
Now we want
to derive the impurity's momentum distribution. Using $\hat{\vec{P}}=\hat{\vec{P}}_{\rm{ph}}+\hat{\vec{P}}_{I}$,
we have
\begin{align}
n_{\rm I}\left(\vec{p}\right) & =\braket{\psi|\delta\left(\hat{\vec{P}}_{I}-\vec{p}\right)|\psi}\\
& =\braket{\psi|\delta\left(\hat{\vec{P}}-\hat{\vec{P}}_{\rm{ph}}-\vec{p}\right)|\psi}\\
& =\braket{\psi|\delta\left(\vec{P}-\hat{\vec{P}}_{\rm{ph}}-\vec{p}\right)|\psi}\\
& =\braket{\psi|\delta\left(\hat{\vec{P}}_{\rm{ph}}-\left[\vec{P}-\vec{p}\right]\right)|\psi}\\
& =n_{\rm ph,\text{tot}}\left(\vec{P}-\vec{p}\right).
\end{align}
In the third line we have replaced $\hat{\vec{P}}\rightarrow\vec{P}$ as
the total momentum $\hat{\vec{P}}$ is fixed for a specific $\ket{\psi}$;
 the Lee-Low-Pines frame block-diagonalizes by total momentum. We
realize that the impurity's momentum distribution $n_{\rm I}\left(\vec{p}\right)$
is given by the phonon momentum distribution $n_{\rm ph,\text{tot}}(\vec{p})$
with the domain of the phonon momentum function mapped from $\vec{p}\rightarrow\vec{P}-\vec{p}$.

\section{Asymptotic Loschmidt Echo Derivation}
\label{appdx:LEchoDeriv}

\begin{figure*}[t!]
	\centering
	\includegraphics[width=0.99\textwidth, 
	height=0.76\textheight]{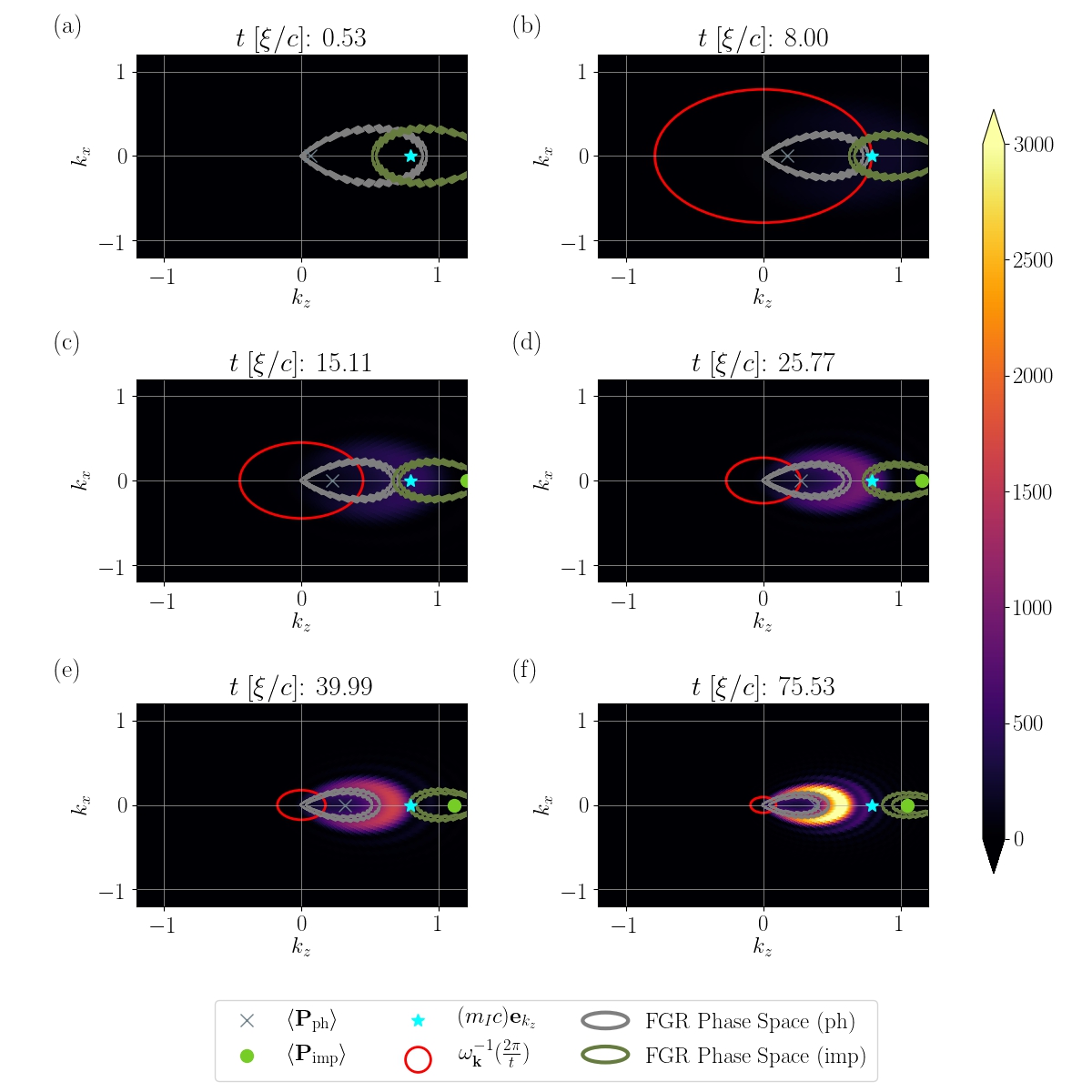}
	\caption{Mean phonon number distribution $n_{\rm ph}(\mathbf{k})$ at different times for a weakly interacting ($a_{\rm IB}^{-1}=-8.92/\xi$) and initially supersonic impurity ($ v_{\rm imp}(t_{0})/c=1.8$). (a) Immediately after the quench we do not build up excitations as enough time hasn't passed. (b) \& (c) We do not build excitations until enough time has passed to excite phonons in the phase space allowed by Fermi's golden rule (FGR). (d) - (f) We build up excitations exactly around this FGR phase space (which contracts over time as the impurity gets slower). The mass ratio is $m_{I}/m_{B}=1$ and we use a sharp UV cutoff $\Lambda=13.91/\xi$.}
	\label{fig:RD_PhononDist}
\end{figure*}

Consider a quench from the FMGS $\ket{0}$ of the non-interacting
state (associated with energy $E_{\downarrow}$) to an interacting
state with dynamics determined by Hamiltonian $\hat{H}_{\uparrow}$.
Let $\left\{ \ket{n_{\uparrow}};E_{\uparrow}^{n}\right\} $ be the
eigenstates and energies of $\hat{H}_{\uparrow}$. The quasiparticle
residue is defined as $Z\equiv\left|\braket{0|0_{\uparrow}}\right|^{2}$
and the Loschmidt echo of a state $\ket{\psi\left(t\right)}$ after
the quench is defined as $S\left(t\right)\equiv e^{iE_{\downarrow}t}\braket{0|\psi\left(t\right)}$.
Inserting the identity $\hat{I}=\sum_{n}\ket{n}\bra{n}$, we have
\begin{align}
S\left(t\right) & =e^{iE_{\downarrow}t}\braket{0|\psi\left(t\right)}\\
& =e^{iE_{\downarrow}t}\braket{0|e^{-i\hat{H}_{\uparrow}t}|0}\\
& =e^{iE_{\downarrow}t}\sum_{n}e^{-iE_{\uparrow}^{n}t}\left|\braket{0|n_{\uparrow}}\right|^{2}\\
& =e^{-i\left(E_{\uparrow}^{0}-E_{\downarrow}\right)t}Z\left(1+\frac{1}{Z}\sum_{n>0}e^{i\left(E_{\uparrow}^{0}-E_{\uparrow}^{n}\right)t}\left|\braket{0|n_{\uparrow}}\right|^{2}\right)\\
& =e^{-i\omega_{\uparrow\downarrow}^{0}t}Z\left(1+\frac{1}{Z}\sum_{n>0}e^{i\omega_{n}t}\left|\braket{0|n_{\uparrow}}\right|^{2}\right)
\end{align}
where we have defined $\omega_{\uparrow\downarrow}^{0}\equiv E_{\uparrow}^{0}-E_{\downarrow}$
and $\omega_{n}\equiv E_{\uparrow}^{0}-E_{\uparrow}^{n}$. We then
have
\begin{align}
\left|S\left(t\right)\right|&=Z\left(1+\frac{1}{Z}\sum_{n>0}e^{i\omega_{n}t}\left|\braket{0|n_{\uparrow}}\right|^{2}\right)\\
&\times\left(1+\frac{1}{Z}\sum_{n>0}e^{-i\omega_{n}t}\left|\braket{0|n_{\uparrow}}\right|^{2}\right)\nonumber
\end{align}
We know that $\omega_{n}<0$ is a monotonic function of $n$ and we
can index the summation and eigenstates $\ket{n_{\uparrow}}$ with
it.
\begin{align}
\left|S\left(t\right)\right|&=Z\left(1+\frac{1}{Z}\sum_{\omega_{n}<0}e^{i\omega_{n}t}f\left(\omega_{n}\right)\right)\\
&\times\left(1+\frac{1}{Z}\sum_{\omega_{n}<0}e^{-i\omega_{n}t}f\left(\omega_{n}\right)\right)\nonumber
\end{align}
where $f\left(\omega_{n}\right)\equiv\left|\braket{0|\omega_{n}}\right|^{2}$.
By the Riemann-Lebesgue lemma, we have
\begin{eqnarray}
\lim_{t\rightarrow\infty}\left|\sum_{\omega_{n}<0}e^{i\omega_{n}t}f\left(\omega_{n}\right)\right|\rightarrow0
\end{eqnarray}
which immediately yields 
\begin{equation}
\lim_{t\rightarrow\infty}\left|S\left(t\right)\right|= Z
\end{equation}
and we see that the Loschmidt echo must asymptote to the quasiparticle
residue at infinite time for all system conditions (initial momentum
and interaction strengths).

\section{Comparison of Dynamics with Fermi's Golden Rule}
\label{appdx:FGR}

As a check of our methods, we examine whether our variational wavefunction agrees with expectations from Fermi's golden rule physics at weak interactions. In Fig. \ref{fig:RD_PhononDist}, we plot the mean phonon number distribution $n_{\rm ph,\text{ind}}(\vec{k})$ at different times for a weakly interacting supersonic impurity; this check gives insight into the power-law decays that were discussed. The green dot (initially off-screen to the right at short times) is the average impurity momentum with the gray 'x' being the average total phonon momentum. The gray lobe in the plots corresponds to the phase space shell of phonons that can be excited according to Fermi's golden rule (FGR) if we limit ourselves to the Fr\"ohlich model. The green lobe corresponds to the accessible momentum shell that the impurity would scatter to after such an emission. The blue star gives a reference for the speed of sound; the impurity stays supersonic so its average impurity momentum is always larger than this.
The region outside the red circle corresponds to the phonon excitations that are accessible to the system at a given time based on the timescale set by their energy. Longer wavelength phonons (closer to $\left|\vec{k}\right|=0$) become accessible at later times.

We see that we excite phonons around the FGR phase space lobe but only after enough time has passed to access those phonons; in the plots this corresponds to the portion of the gray lobe that is outside the red circle. As time elapses, the impurity slows down towards the speed of sound and the FGR phase space lobe therefore contracts; we thus see phonons excitations build up in the entire area swept out by the FGR lobe.

\end{document}